\documentclass[aps,prl,twocolumn,groupedaddress,showpacs,floatfix]{revtex4}

\usepackage{graphicx,color}
\usepackage{amssymb}
\usepackage{amsbsy}

\usepackage{amsmath}

\usepackage{cases}

\usepackage[overload]{empheq}

\begin{document}


\title{Fraction of clogging configurations sampled by granular hopper flow}


\author{C. C. Thomas and D. J. Durian}
\affiliation{Department of Physics and Astronomy, University of Pennsylvania, Philadelphia, PA 19104-6396, USA}


\date{\today}

\begin{abstract}
We measure the fraction $F$ of flowing grain configurations that precede a clog, based on the average mass  discharged between clogging events for various aperture  geometries. By tilting the hopper, we demonstrate that $F$ is a function of the hole area projected in the direction of the exiting grain velocity.  By varying the length of slits, we demonstrate that grains clog in the same manner as if they were flowing out of a set of smaller independent circular openings.  The collapsed data for $F$ can be fit to a decay that is exponential in hole width raised to the power of the system dimensionality.  This is consistent with a simple model in which individual grains near the hole have a large but constant probability to precede a clog.  Such a picture implies that there is no sharp clogging transition, and that all hoppers have a nonzero probability to clog.
\end{abstract}

\pacs{45.70.Ht, 45.70.Mg}

\maketitle

Granular flow through a small hole or bottleneck is susceptible to clogging, where a stable arch (or dome) forms over the hole and arrests the entire flow \cite{To01, Zuriguel03, Zuriguel05}.  Clogging is highly useful for probing the extreme limits of granular flow, since the bulk continuum-like behavior of a large system is governed by a relatively small number of particles at the exit. Furthermore, clogging is a natural phenomenon that illustrates spontaneous evolution from a freely-flowing state to a jammed state with no change in the external forcing. Similar issues are important for understanding the flow of suspensions \cite{Valdes06, Roussel07,Chen12} through constrictions, the flow of vortices through an array of pinning sites in superconductors \cite{Reichhardt10, Reichhardt12}, as well as automotive \cite{CarTraffic} and pedestrian \cite{Helbing00} traffic.  In spite of many simulations \cite{Manna00, Pournin07, Tewari13, Kondic14} and experiments in both two- \cite{Clement, To01, To05, Janda08, Tang09, Obstacle} and three-dimensional hoppers \cite{Drescher, Zuriguel03, Zuriguel05, Pournin07, Sheldon, Saraf, CloggingPD}, the ability to predict or control clogging is still lacking \cite{ZuriguelPIP14}.

\begin{figure}
\includegraphics[width=3 in]{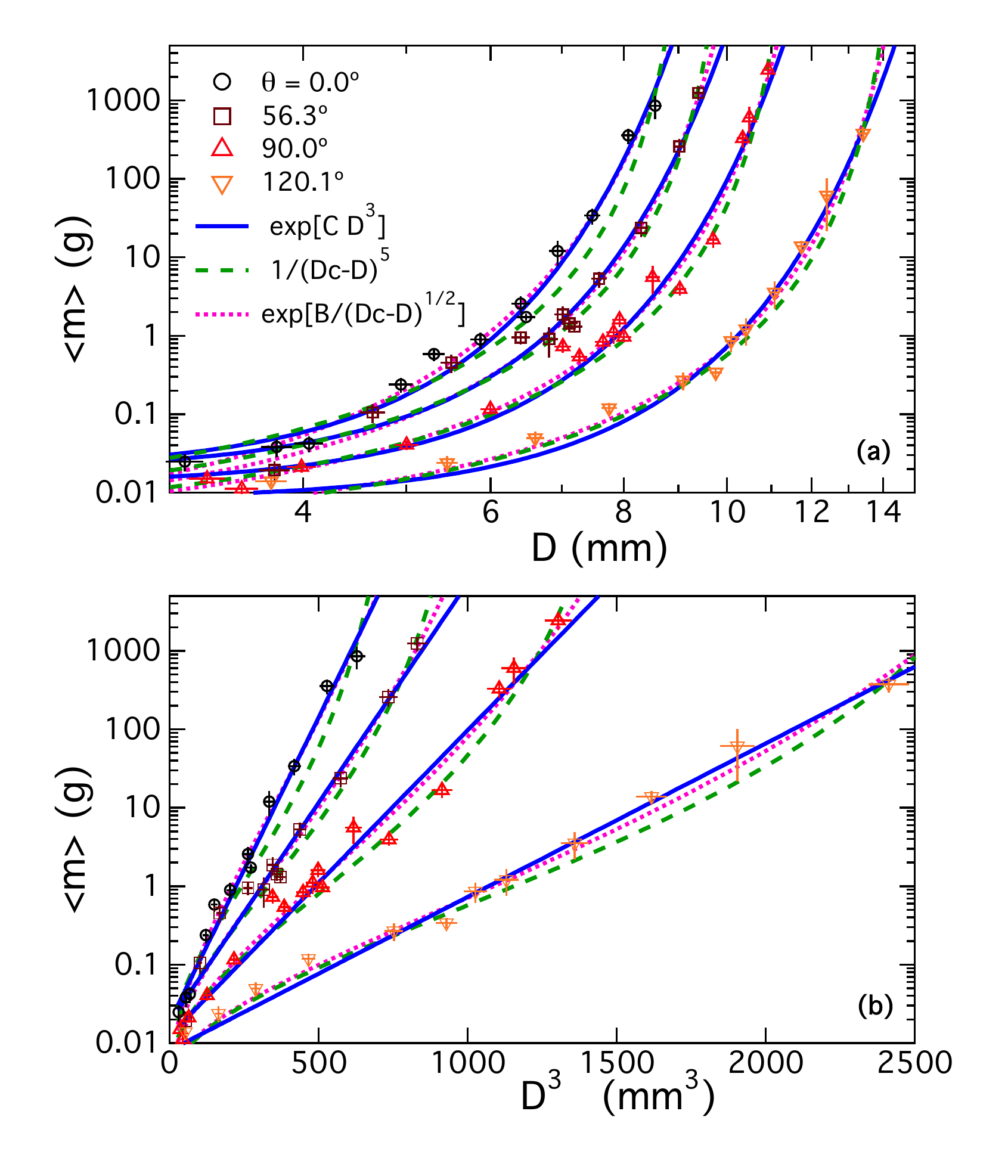}
\caption{(Color online) Average mass $\langle m \rangle$ discharged before a clog forms, illustrated as a function of (a) hole diameter $D$ and (b) $D^3$ for $d=2$~mm glass beads in hoppers with different tilt angles, $\theta$.  These data are from Ref.~\cite{CloggingPD}, and are well fit by a variety of functions as labeled. Vertical error bars are based on measured $m$ values for 10 or more discharge events; horizontal error bars reflect uncertainty in $D$.}
\label{MassD}
\end{figure}

One measure commonly used to quantify clogging is the average mass $\langle m \rangle$ discharged before a clog occurs.  Fig.~\ref{MassD} displays data from Ref.~\cite{CloggingPD}, showing how $\langle m \rangle$ grows extremely rapidly with hopper hole diameter $D$ for several different tilt angles $\theta$.
Various forms for $\langle m \rangle$ versus $D$ have been proposed. Refs.~\cite{Zuriguel05, Janda08, CloggingPD} fit the data to a critical power law (long dashed curves on Fig.~\ref{MassD}), diverging at finite $D = D_c$. However, there is no model for the value of $D_c$ or for the unusually large exponents, which are in the range of $5-12$.   Furthermore, other data show that $\langle m \rangle$ grows in proportion to $\exp(C D^2)$ in two-dimensional systems \cite{To05, Janda08} or $\exp(C D^3)$ in a three-dimensional system (solid curves on Fig.~\ref{MassD}) \cite{CloggingPD}.   The exponential form is somewhat better in describing the Ref.~\cite{CloggingPD} data, since the ratio of $\chi^2$ for the diverging form to its value for the exponential form falls mainly in the range $0.7-1.9$.
The conflict between these different forms is significant.  If $\langle m \rangle$ diverges at finite $D$, then there exists a sharp clogging transition marking a regime where the flow will \emph{never} clog.

To clarify the nature of clogging, we propose an alternative quantity to $\langle m\rangle$: the fraction $F$ of grain configurations near the exit that cause the flow to clog. We argue how $F$ may be deduced from $\langle m \rangle$, using the Poisson character of the clogging process, without any knowledge whatsoever required about the actual grain microstates.  By tilting the hopper, and by varying the length $L$ of a slit, we obtain a range of data that can be collapsed together using a probabilistic argument onto a single curve for $F$ versus hole diameter.  The results are successfully modeled based on single-grain behavior in a volume near the aperture, where each grain has a constant probability to clog independent of hole size.

\begin{table}
\caption{\label{tab:Materials}The mean grain diameter $d$, the bulk density $\rho$, and the draining angle of repose $\theta_r$ for the grains tested. Errors for $d$  indicate the standard deviations of the distributions. For the other material properties, error bars indicate the standard error in the measurements.  The error bars plotted in all figures reflect the propagated uncertainty in grain properties, uncertainties in hole diameter (ranging from 0.013 to 0.2~mm in each dimension), uncertainty of $\pm0.5^\circ$ in tilt angle, and uncertainty in $\langle m\rangle$ based on $m$ values for 10 or more discharge events.}
\begin{ruledtabular}
\begin{tabular}{lllll}
	\multicolumn{1}{c}{Material} & \multicolumn{1}{c}{$d$ (mm)} & \multicolumn{1}{c}{$\rho$ (g/cm$^3$)} & \multicolumn{1}{c}{$\theta_r$} \\ \hline
	Glass Spheres & $2.02 \pm 0.04$ & $1.62 \pm 0.01$ & $23.5\pm 0.5^{\circ}$\\
	Glass Spheres & $0.96 \pm 0.05$  &$1.56 \pm 0.02$ & $20.0\pm0.4^{\circ}$ \\
	Dry Tapioca & $3.50 \pm 0.14$  & $0.69 \pm 0.01$ & $32.3 \pm 0.5^{\circ}$ \
\label{Materials}
\end{tabular}
\end{ruledtabular}
\end{table}

Our experiments utilize three different spherical grains and three different hopper shapes. The material properties of the grains, including the grain diameter $d$, bulk density $\rho$, and draining angle of repose $\theta_r$, are listed in Table~\ref{Materials}.  For clogging of grains at a circular hole, we use an aluminum hopper with inner cross-section of dimensions $9.5\times 9.5$~cm$^2$. The hole is a camera iris with continuously adjustable diameter $D$. We also change the propensity to clog by tilting this hopper, varying the angle $\theta$ that the hole makes with gravity. When $\theta > 60^\circ$, the iris is mounted on the sidewall of the hopper rather than its floor; however, this does not affect the behavior of the flow or clogging. Data for 2~mm glass spheres in this hopper were previously published in Ref.~\cite{CloggingPD}.

To explore the effect of the aperture shape, we investigate clogging from a rectangular slit of long dimension $L$ and short dimension $D$. As with the circular hole, we use a hopper with smooth horizontal floor and vertical sidewalls.   The inner cross-section of this hopper is  $28\times 20$~cm$^2$. To vary $D$, we use a custom-made aluminum slit that allows for fine control over the width while maintaining a constant length. Data collected for the 1~mm glass spheres for this slit with $L$ = 149~mm were previously reported in Ref.~\cite{CloggingPD}. However, we now also change the slit length by masking the sides of the slit. We investigate the effect of changing $L$ and $D$ using both the 1~mm and 2~mm glass spheres.

Finally, we report on clogging for a somewhat different slit geometry and material.  This third hopper has a slit of constant length $L$ = 3.8~cm and adjustable width $D$. However, the inner dimensions of the hopper are  $3.8~\times~56$~cm$^2$ and thus two of the inner walls coincide with the ends of the slit.  Furthermore, we inhibit the sliding of grains along the bottom of the hopper by placing two grain-size dowels at the lips of the slit. For this hopper we use 3.5~mm diameter dry tapioca pearls.

To determine the fraction $F$ of flowing configurations that precede a clog, we begin by noting that the distribution of flow durations $\tau$ is widely regarded as exponential: $P\left(\tau\right) \propto~\exp(-\tau/\langle \tau \rangle)$, where $\langle \tau \rangle$ is the average duration of flow events \cite{Clement, Zuriguel03, Zuriguel05, To05, Janda08, Tang09, CloggingPD, Obstacle}.  Clogging is thus a Poisson process, where the probability to remain unclogged across a small time increment is $1-{\rm d}t/\langle \tau \rangle$.  Physically, we may picture that flow brings new grains into the aperture region, and different flow configurations are successively sampled at random, until one arises that precedes a clog.  Three ingredients are now needed.  First, grains moving down at speed $v$ experience a new configuration --with new opportunity to clog-- when they have moved a distance $\ell=v\tau_0$, where $\tau_0$ is the configuration lifetime and $\ell$ is a corresponding sampling length anticipated to be roughly one grain size.  Second, the ratio $\tau_0/\langle \tau \rangle$ is recognized as the fraction $F$ of configurations that precede formation of a stable arch over the aperture, since the number of configurations sampled in the average discharge event is $\langle \tau\rangle/\tau_0$.  Third, the average discharge mass is $\langle m \rangle = \rho A v \langle \tau \rangle$, where $\rho$ is the bulk mass density of the medium and $A$ is the aperture area.  Altogether, these three ingredients give the fraction $F$ of pre-clogging flow configurations as
\begin{equation}
\label{FEq}
	F = \rho A \ell/\langle m \rangle.
\end{equation}
It is remarkable and powerful that $F$ may thus be deduced from the right-hand side without actually measuring grain positions, orientations, momenta, or contact forces.  The only proviso is that at each time increment the clogging probability be constant; e.g. there should be no transients from using an overly-large impulse to start the flow.  This is supported by the continued Beverloo behavior for discharge rates below clogging in Fig.~1 of Ref.~\cite{CloggingPD}, as well as by the discussion in \cite{supp}.

\begin{figure}
\includegraphics[width=3 in]{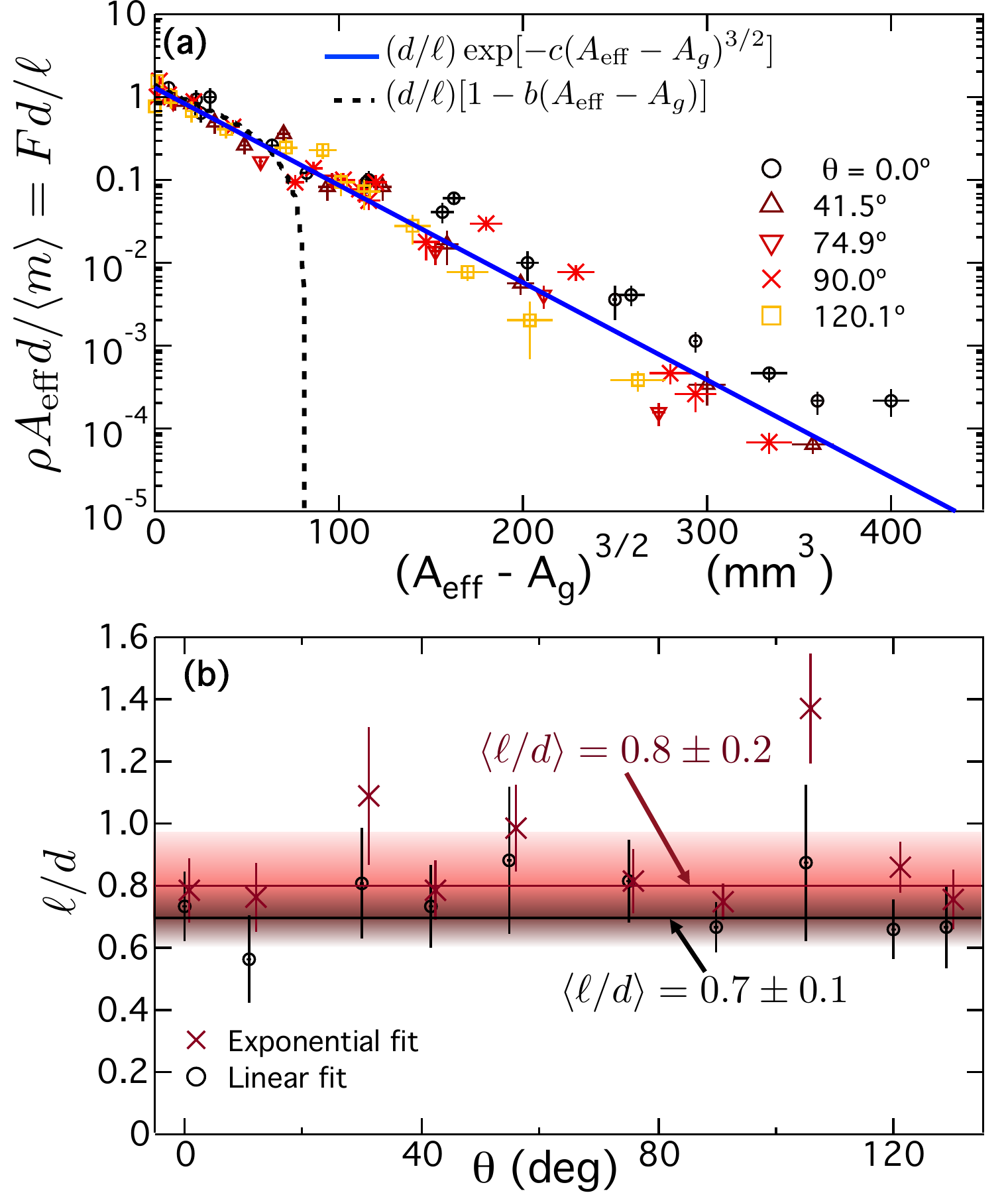}
\caption{(Color online)  (a) $\rho A_{\rm eff} d/\langle m \rangle$ versus $(A_{\rm eff} - A_g)^{3/2}$ for tilted hoppers, where $\rho$ is the bulk density of the grains, $d=2$~mm is the grain diameter, $\langle m \rangle$ is the average mass discharged before clogging, $A_{\rm eff}$ is the effective hole area for a tilted hopper, and $A_g$ is the cross-sectional grain area. According to Eq.~(\ref{FEq}), the quantity plotted is equal to $F d/\ell$, where $\ell$ is the sampling length. (b) Sampling length $\ell/d$ versus tilt angle $\theta$, as found from fits as displayed in part (a). The average values of $\langle \ell\rangle$ for linear and exponential fits are displayed; combining these, we henceforth take the sampling length as $\ell = (0.75\pm0.20)d$.}
\label{GetEllBoth}
\end{figure}

In order to estimate the sampling length $\ell$, we assume $F$ approaches one as the aperture area $A$ is reduced to the size of one grain. We plot $\rho A d/\langle m \rangle$ versus $(A - A_g)^{3/2}$ in Fig.~\ref{GetEllBoth}a, where $A_g$ is the cross-sectional area of a single grain. We do this for the 2~mm glass spheres clogging at a circular hole. We also include data for tilted hoppers, where $A$ is replaced by an effective area defined by the projection of the hole $A \boldsymbol{\hat{n}}$ in the average flow direction $\boldsymbol{\hat{v}}$:
\begin{subnumcases}{A_{\rm eff} \equiv \left(A \boldsymbol{\hat{n}}\right) \cdot \boldsymbol{\hat{v}} = A \label{AeffEq}} 
    \!\! \cos \theta & $\theta \leq \theta_r$ \quad \quad  \label{eq2a} \\
    \!\!  \cos \left[\left(\theta+\theta_r\right)/2 \right]  & $\theta \geq \theta_r$ \quad \label{eq2b}
\end{subnumcases}
where $\theta_r$ is the draining angle of repose (see Table~\ref{Materials}) and $\theta$ is the angle between the hole normal $\boldsymbol{\hat{n}}$ and vertical \cite{CloggingPD}.  Fig.~\ref{GetEllBoth}a demonstrates two different fitting forms that may be used to estimate the $y$-intercept, which we take as $d/\ell$.  The results displayed Fig.~\ref{GetEllBoth}b show no trend with $\theta$ and are roughly the same for the two fitting forms.   Averaging over $\theta$ gives a combined estimate of the sampling length that is slightly less than one grain diameter, as expected: $\ell = \left(0.75 \pm 0.20\right) d$.  This result is consistent with an alternative measurement in Ref.~\cite{supp}.

\begin{figure}
\includegraphics[width=3 in]{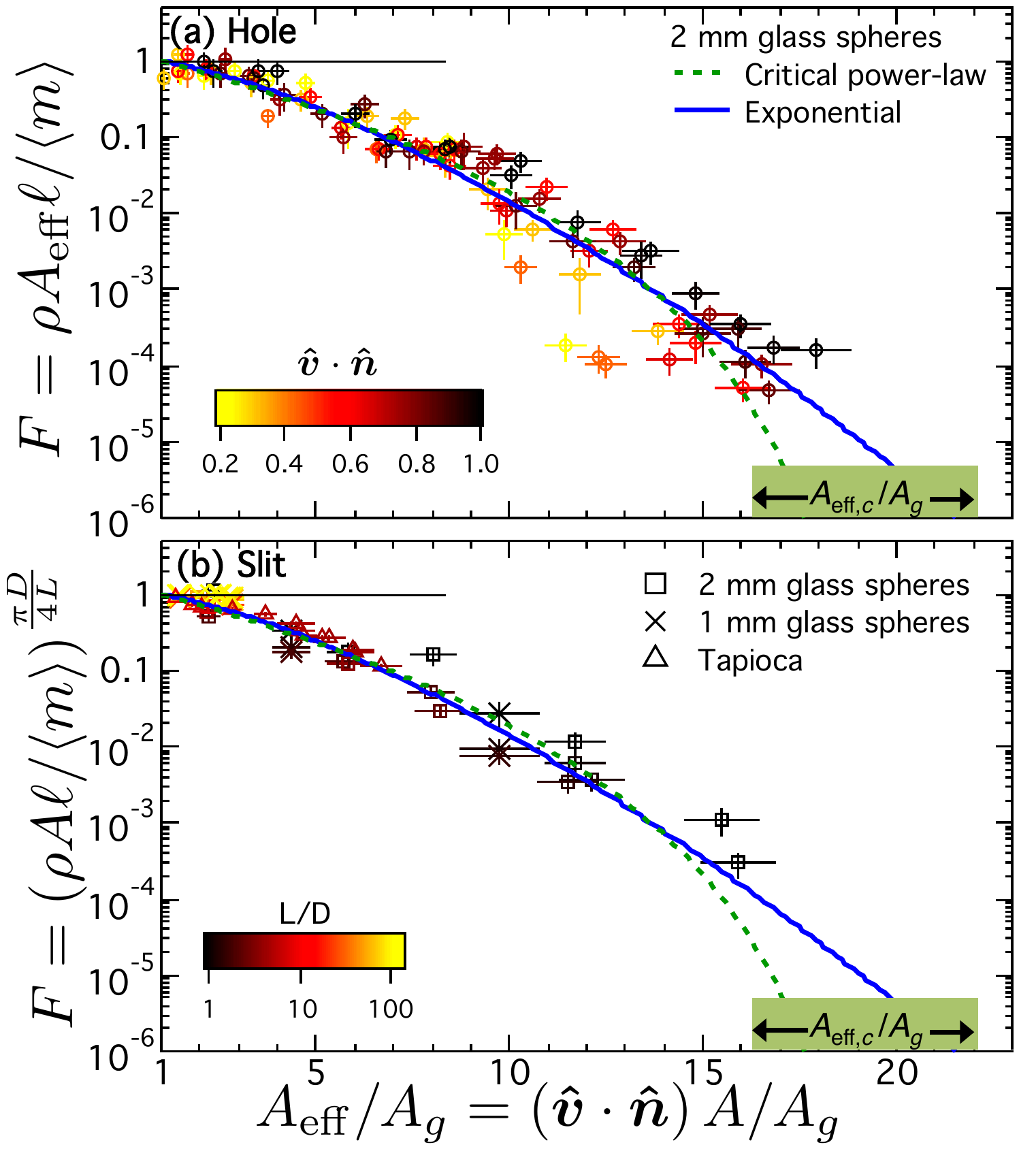}
\caption{(Color online) Fraction of  configurations that precede a clog, deduced from measurements of $\langle m \rangle$, versus normalized effective hole area.
In (a), data points for circular holes are shaded by $\boldsymbol{\hat{v}} \cdot \boldsymbol{\hat{n}}$.  In (b), data points for hole-equivalent subsections of rectangular slits are shaded by slit aspect ratio $L/D$.  In both the grain sizes are indicated in the legend.  The clogging transition locations found from critical power law fits in Ref.~\cite{CloggingPD} for the 2~mm glass spheres are indicated by $A_{\rm eff,c}/A_g$ at bottom right. The best fits to Eq.~(\ref{FPow}) and Eq.~(\ref{FExp}) are overlaid as dashed and solid curves, respectively.}
\label{FBoth}
\end{figure}

We now find $F$ for the 2~mm glass spheres in the tilted hopper, using $\ell/d=0.75$ with the measured average discharged mass $\langle m \rangle$ for circular holes of different diameters $D$ and tilt angles $\theta$.  Fig.~\ref{FBoth}a shows $F$ as a function of the dimensionless effective hole area $A_{\rm eff}/A_g$. The data points in this plot are shaded by $\boldsymbol{\hat{v}} \cdot \boldsymbol{\hat{n}}$. These values of $F$ for circular holes fall off rapidly by many orders of magnitude with increasing $A_{\rm eff}/A_g$. Furthermore, there is apparently no dependence of $F$ on $\theta$, such that $F$ is a single function of $A_{\rm eff}/A_g$.

Next we consider rectangular slits of long dimension $L$ and short dimension $D$, at $\theta = 0$. We hypothesize that slits behave as a collection of $\left(4/\pi\right) L/D$ independently clogging circular holes of diameter $D$, giving
\begin{equation}
	F_{\rm slit}\left(D,L\right) = \left[F_{\rm circle}\left(D\right)\right]^{\left(4/\pi\right) L/D}.
\label{SlitEq}
\end{equation}
We therefore determine the value of $F$ for circular holes of diameter $D$, using measured values of $\langle m \rangle$ for rectangular slits of width $D$ and again taking $\ell/d=0.75$.  These data are displayed on Fig.~\ref{FBoth}b versus $A~=~\pi D^2/4$, where $D$ is the slit width and the aspect ratio $L/D$ is indicated by shading.  This analysis causes the data in Fig.~\ref{FBoth}b, for the different grain types and a wide range of slit widths and aspect ratios, to all fall onto a single curve.  This collapse supports the hypothesis of Eq.~(\ref{SlitEq}).

Finally, we compare $F$ for circular holes (Fig.~\ref{FBoth}a) with the effective circular components constituting the rectangular slits (Fig.~\ref{FBoth}b) by simultaneously fitting both data sets to two empirical forms:
\label{FFits}
  \begin{align}[left ={F(x) = \empheqlbrace}]
    & \left(\frac{{x_c}^\gamma - x^\gamma}{{x_c}^\gamma - 1}\right)^{\beta}\label{FPow}\\
    & \exp\left[-C\left(x^{\alpha/2}-1\right)\right]\label{FExp}   , 
   \end{align}
where $x \equiv A_{\rm eff}/A_g$. For the best critical power-law fit to both hole and slit data, we find $\gamma = 1.0 \pm 0.2$, $\beta = 6 \pm 1$, and $x_c = 19.5 \pm 1.5$.   This fit is shown by the dashed curve in Fig.~\ref{FBoth}a and \ref{FBoth}b.  Note that the critical aperture size $x_c$ is within the range of the observed values for the clogging transition reported in Ref.~\cite{CloggingPD}, indicated on the lower right of Fig.~\ref{FBoth}.   For the best simultaneous exponential fit to both hole and slit data, we find $\alpha = 3.0 \pm 0.2$ and $C = 0.14 \pm 0.03$.  This result is shown by the solid curve in Fig.~\ref{FBoth}a and \ref{FBoth}b.  Note that the two fitting forms describe both data sets equally well, as the ratio of $\chi^2$ for the fit to Eq.~(\ref{FPow}) to its value for the fit to Eq.~(\ref{FExp}) is 1.07. Thus, the data collapse onto a single function of $A_{\rm eff}/A_g$.  However, as in Fig.~\ref{MassD}, the multitude of good empirical fitting forms makes it unclear whether or not there is a sharp clogging transition where $F(x)$ vanishes at finite $x_c$.

It is possible to justify Eq.~(\ref{FExp}), but not Eq.~(\ref{FPow}),  with a na\"ive model based on the possible microstates of individual grains near the aperture.  To begin we suppose that the number of grains in the clogging region above the hole must scale as $N=(D/d)^\alpha$, where $\alpha$ depends on dimensionality.  If each of these $N$ grains can be in $V_1$ single-grain position / momentum / contact-force microstates, then the total number of allowed configurations is $\Omega \propto {V_1}^N$ (this is an uncontrolled approximation in ignoring collective effects).  If only a certain number $V_{pc}$ of these single-grain microstates precede a clog, then the total number of clogging configurations is $\Omega_c \propto {V_{pc}}^N$.  And then from $F \equiv \Omega_c/\Omega$ we find
\begin{equation}
	F = \left(\frac{V_{pc}}{V_1}\right)^N =  \exp   \left[-\left(\ln \frac{V_1}{V_{pc}}\right)\left(\frac{D}{d}\right)^\alpha\right].
\label{Fvsvf}
\end{equation}
This is precisely Eq.~(\ref{FExp}) for large holes, if $V_{pc}/V_1$ is constant and if $\alpha=3.0\pm0.2$.  For our data in three dimensions, and also for the $\langle m \rangle \propto \exp(CD^2)$ data in two dimensions \cite{To05, Janda08}, we thus have a consistent picture in which the fraction of clogged configurations dies exponentially with hole width raised to a power $\alpha$ equal to dimensionality.  The value of $\alpha$ implies that more grains are involved in arch formation than just those exposed underneath in the final static monolayer spanning the hole.  This agrees with observations that an obstacle placed over the aperture has no influence if it is higher than about two hole widths \cite{Obstacle}.

\begin{figure}
\includegraphics[width=3 in]{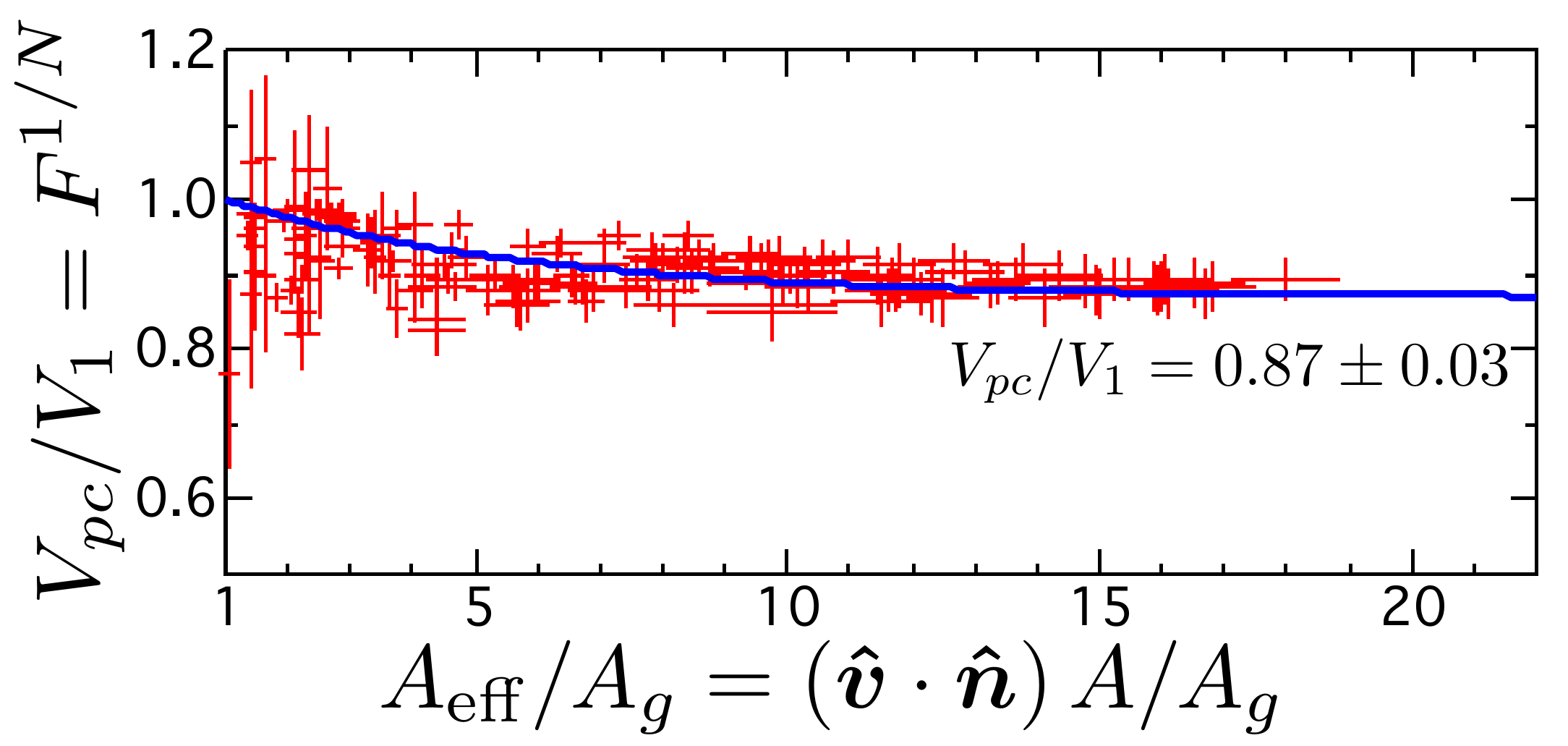}
\caption{(Color online) Ratio of the stable to total single-grain microstates versus normalized effective hole area, obtained for \emph{all} the data shown in Fig.~\ref{FBoth} by computing $F^{1/N}$ where $N = \left(\boldsymbol{\hat{v}} \cdot \boldsymbol{\hat{n}}\right)(D/d)^3$. For moderately large holes, $V_{pc}/V_1$ falls to a constant value $0.87 \pm 0.03$, as illustrated by the exponential fit shown in blue.}
\label{vsvf}
\end{figure}

The above model for $F$ is highly oversimplified, but can be pushed for an estimate the ratio $V_{pc}/V_1$ of single-grain flowing microstates near the hole that precede a clog.  First, for large holes, comparison of Eq.~(\ref{Fvsvf}) and the fit to Eq.~(\ref{FExp}) gives $V_{pc}/V_1=\exp(-C)=0.87\pm0.03$.  More generally, we can deduce the ratio from $V_{pc}/V_1=F^{1/N}$ using $N=\left(\boldsymbol{\hat{v}} \cdot \boldsymbol{\hat{n}}\right)(D/d)^3$.  The results from all hole and slit data are plotted in Fig.~\ref{vsvf}.  For small hole sizes, $V_{pc}/V_1$ approaches 1.  With increasing hole size, $V_{pc}/V_1$ does not vanish but instead asymptotes to a nonzero constant $0.87\pm0.03$.  This value is perhaps surprisingly large.  It means that each grain near the exit is almost always in a position to participate in clog formation, which is consistent with the flows being very dense and not far dilated from random-close packing.  For large enough hole sizes, actual clog formation is rare because \emph{all} the grains in the region must be suitably positioned.  Perhaps $V_{pc}/V_1$ represents the ratio of stable- to free- volume, or perhaps momenta and contact forces are important too.  The detailed nature of the single-grain microstates, the size of the clogging region, the role of dissipation, and the importance of collective effects could all be profitably studied by simulation and perhaps even two-dimensional experiments.

To conclude, prior fits to critical power laws \cite{Zuriguel05, Janda08,  CloggingPD} and the seeming universality of the clogging phase diagram \cite{CloggingPD} were strong arguments for the existence of a sharp clogging transition at a finite hole size $D_c$.  However, lack of theory for exponents and for maximum hole size, and the absence of diverging quantities on approach to $D_c$ from above, were all causes for concern.  Here, based on a probabilistic analysis and modeling of new and old data, we find compelling reasons to view clogging as always possible.  The act of flow causes a random sampling of configurations for grains above the hole in a region of size given by hole width raised to the power of dimensionality.  Flow proceeds until a configuration arises in which \emph{all} these grains are able to participate in forming a stable arch, or dome.
This naturally gives an exponential distribution of discharge event sizes.
And since the probability for a given grain to be able to participate in a clog is constant, independent of hole size, there is no critical hole size beyond which the system is no longer susceptible to clogging.  Instead, the average flow duration grows dramatically, exponentially, due to the increasingly large number of grains that are required to occupy pre-clogging flow microstates.  Clogging can thus be seen as a phenomenon similar to the glass and jamming transitions, which are defined by an observation threshold.  There, relaxation times grow dramatically and can be fit to a variety of forms, some of which diverge and some of which do not.  The same can be said for clogging.  There is no critical point, but as a practical matter a clogging transition may still be defined at the hole size beyond which clogging becomes so vastly improbable as to be essentially unobservable.

\begin{acknowledgments}
This work was supported by the NSF through Grant No. DMR-1305199. We thank R. P. Behringer and E. Cl\'ement for helpful discussions and for raising the possibility that $\langle m \rangle$ does not diverge at finite hole size.
\end{acknowledgments}

\bibliography{FofA_Refs}

\end{document}